\documentstyle[psfig,aps]{revtex}
\begin{document}
\draft

\title{Scaling of curvature in sub-critical gravitational collapse}

\author{David Garfinkle
\thanks {Email: garfinkl@vela.acs.oakland.edu}}
\address{
\centerline{Department of Physics, Oakland University,
Rochester, Michigan 48309}}

\author{G. Comer Duncan
\thanks {Email: gcd@chandra.bgsu.edu}}
\address{
\centerline{Department of Physics and Astronomy, Bowling Green State University,
Bowling Green, Ohio 43403}}

\maketitle

\null\vspace{-1.75mm}

\begin{abstract}

We perform  numerical simulations of the gravitational collapse of a spherically
symmetric scalar field.  For those data that just barely do not form black holes 
we find the maximum curvature at the position of the central observer.  We find a 
scaling relation between this maximum
curvature and distance from the critical solution.  The scaling relation is
analogous to that found by Choptuik for black hole mass for those data that do
collapse to form black holes.  We also find a periodic wiggle in the scaling
exponent.
 
\end{abstract}

\pacs{PACS 04.20.-q, 04.20.Fy, 04.40.-b}

\section{Introduction}

Choptuik has found scaling phenomena in gravitational 
collapse.\cite{choptuik}
He numerically evolves a one parameter family of initial data for a spherically
symmetric scalar field coupled to gravity.  Some of the data collapse to form
black holes while others do not.  There is a critical value of the parameter
separating those data that form black holes from those that do not.  The
critical solution (the one corresponding to the critical parameter) has the
property of periodic self similarity: after a certain amount of logarithmic
time the profile of the scalar field repeats itself with its spatial scale
shrunk.  For parameters slightly above the critical parameter the mass of
the black hole formed scales like $ {(p-p*)}^\gamma$ where $p$ is the parameter, 
$p*$ is its critical value and $\gamma$ is a universal scaling
exponent that does not depend on which family is being evolved.  
Numerical simulations of the critical gravitational collapse of other types of
spherically symmetric matter were subsequently performed.  
These include complex scalar fields \cite{eardley}, perfect fluids \cite{evans1},
axions and dilatons \cite{stewart}, and Yang-Mills fields \cite{bizon}.  
In addition scaling has been found in the collapse of axisymmetric gravity 
waves \cite{evans2}, and in a perturbative analysis of fluid collapse with no
symmetries\cite{gundf}.  Thus scaling seems to be a generic feature of critical gravitational
collapse. In some of these systems the critical solution has periodic self-similarity
while in other systems it has exact self-similarity.  

These phenomena were discovered numerically, so one would like to have an
analytic explanation for why systems that just barely undergo gravitational
collapse behave in this way.  An explanation of the scaling of black hole mass
was provided by Koike, Hara and Adachi \cite{hara}.  These authors assume  
that the critical solution is exactly self-similar and has exactly one unstable
mode.  Exact self-similarity means that the critical solution has a homothetic
Killing vector, {\it i.e.} a vector field $\xi$ such that ${{\cal L}_{\xi }}
{g_{ab}} = - \, 2 \, {g_{ab}}$.  Let coordinates be chosen so that $ {\xi } = \partial /
\partial t$.  Then the unstable mode grows as $ e^{\kappa t} $ for some constant
$\kappa $.  The result of reference\cite{hara} is that black hole mass scales as
$ {(p-p*)}^\gamma$ where $ \gamma = 1/\kappa $.  

The results of reference\cite{hara} were extended to the case of periodic
self-similarity by Gundlach\cite{gundlach1} and by Hod and Piran \cite{piran2}.  
Here, the assumptions are
that  the critical solution is periodically self-similar and has exactly one unstable
mode.  Periodic self-similarity means that there is a diffeomorphism $\zeta $
and a number $\Delta $ such that $ {\zeta ^*} ( {g_{ab}} ) = {e^{ - 2 \Delta }}
\, {g_{ab}}$.  Let coordinates be chosen so that $\zeta $ is the transformation
$ t \to t + \Delta $ with the other coordinates remaining constant.  Then the
unstable mode grows as $ e^{\kappa t} $ multiplied by a function that is
periodic in $t$.  (Here again $\kappa $ is a constant).  The result is still a
scaling relation for black hole mass; but it is more complicated than a linear
relation.  A graph of $\ln M$ vs. $ \ln (p - p*) $ is no longer a straight
line; but is instead the sum of a linear function and a periodic function.  The
slope of the linear function is again $\gamma = 1/\kappa $ and the period of
the periodic function is $\Delta /(2 \gamma )$.  The scaling relation for
black hole mass in scalar field collapse was originally thought to be
linear\cite{choptuik}  because the additonal ``wiggle'' is small.  This small
wiggle was found numerically by Hod and Piran\cite{piran2}.  

Any proposed analytic explanation of a numerically observed phenomenon needs to be
tested.  Perhaps the best such test is to ask what other phenomena are predicted
by the explanation and then to see whether those phenomena occur.  
One remarkable property of the derivations in references\cite{hara,gundlach1,piran2} is
their generality.  The only property of black hole mass that is used is that it
is a global property of the spacetime and has dimensions of length.  Furthermore,
the derivations apply as well to solutions that do not collapse to form
black holes as to those that do: the only assumption needed  
is that the initial data be near data that leads to the critical solution.

Therefore, it is a consequence of the explanation of 
references\cite{hara,gundlach1,piran2} that other
scaling relations exist in near-critical collapse, even in the case 
where no black hole forms.  This is in contrast to the case of phase transitions,
where scaling behavior occurs only on one side of the critical parameter. 
For the case of a periodically self-similar critical solution, the
scaling relation should have a wiggle with period $\Delta /(2 \gamma )$.  For
the case where no black hole forms, the field collapses for a while and then
disperses.  Therefore, at the position of the central observer, the scalar
curvature should grow, achieve some maximum value $R_{\rm max}$ and then
approach zero at late times.  This  $R_{\rm max}$ is a characteristic of the
spacetime and has dimensions of $ {\rm length}^{- 2} $.  Therefore, one would
expect that a graph of $\ln {R_{\rm max}} $ vs. $\ln (p* - p) $ should be a
curve with average slope $ - \, 2 \gamma $ (where $\gamma $ is the same
constant that occurs in the black hole mass scaling relation) and a wiggle
with period $\Delta /(2 \gamma )$.  

In order to test the explanation of references\cite{hara,gundlach1,piran2},
this paper presents the results of  numerical investigations which 
explore whether
$R_{\rm max}$ obeys exactly this sort of scaling relation.  We have performed 
numerical simulations of the collapse of a family of initial data for 
a spherically
symmetric scalar field.  The data were chosen to be near the critical solution,
but with $p < p*$ so that no black hole forms.  For each evolved spacetime in
the family we find $R_{\rm max}$, the maximum scalar curvature at the
central observer.  We then plot $\ln {R_{\rm max}}$ vs 
$ \ln (
p* - p)$ and show that the resulting curve is a straight line with 
a periodic wiggle,
where the slope of the line is $ - \, 2 \gamma $ and the period of the wiggle
is $\Delta /(2 \gamma )$.  Section II briefly presents the 
numerical method used.
The results are presented in Section III.  Section IV contains a 
discussion of some
of the implications of the results of these studies.

\section{Numerical method}

The numerical method used is that of Garfinkle\cite{garfinkle}.  This method is
a modification of an earlier method due to Goldwirth and Piran\cite{piran1},
which is in turn based on analytical work by Christodoulou \cite{christodoulou}.
The matter is a massless, minimally coupled scalar field, 
with both scalar field and
metric spherically symmetric.  In addition to the usual area coordinate $r$ and
angular coordinates, we use a null coordinate $u$ defined
to be constant along outgoing light rays, and equal to proper time of the
central observer at $r=0$.  Instead of directly using the matter field
$\phi
$, it is convenient to work with the quantity $ h \equiv (\partial / \partial
r) ( r \phi )$.  Due to the spherical symmetry, the metric is completely
determined by the matter.  This is made explicit as follows: for any function
$f(u,r) $ define
$$
{\bar f} (u,r) \equiv {1 \over r} \; {\int _0 ^r} \; f(u,{\tilde r}) \; d
{\tilde r} \; \; \; .
\eqno(1)
$$
Then define the quantity $g$ by 
$$
g(u,r) \equiv \exp \left ( 4 \pi \; {\int _0 ^r} \; 
{{d {\tilde r}} \over {\tilde
r}}
\; {{\left [ h(u, {\tilde r}) - {\bar h} (u, {\tilde r})\right ] }^2} \right )
\; \; \; .
\eqno(2)
$$
Then the metric is given by
$$
d {s^2} = - \, g \, d u \, ( {\bar g} \, d u \, + \, 2 \,  d r) 
\; + \; {r^2} \, d {\Omega
^2} 
\eqno(3)
$$
where $ d {\Omega ^2} $ is the usual unit two-sphere metric.  
Define the operator
$$
D \equiv {\partial \over {\partial u}} \; - \; {1 \over 2} \; {\bar g} \;
{\partial \over {\partial r}} \; \; \; .
\eqno(4)
$$
$D$ is essentially the derivative operator along ingoing light rays.  
The evolution
equation for $h$ is 
$$
D \, h = {1 \over {2 r}} \; (g - {\bar g}) (h - {\bar h}) \; \; \; .
\eqno(5)
$$

The numerical treatment of these equations is as follows: given $h$ at some time
$u$, the quantities ${\bar h},\, g$ and $\bar g$ are evaluated in turn, with the
integrations done using Simpson's rule.  This is quite accurate, except near
$r=0$.  For the region near $r=0$ we use a Taylor series method: first we fit
the values of $h$ near $r=0$ to a straight line and thus evaluate the quantity
$$
{h_1} \equiv {{\left . {{\partial h} \over {\partial r}} \right | }_{r=0}} \; \;
\; .
\eqno(6)
$$
Near $r=0$ the quantities $ {\bar h} , \, g $ and $\bar g $ are then given by
$$
{\bar h} = h \; - \; {1 \over 2} \; {h_1} \, r \; + \; O ( {r^2} ) \; \; \; ,
\eqno(7)
$$
$$
g = 1 \; + \; {\pi \over 2} \; {h _1 ^2} \, {r^2} \; + \; O ( {r^3} ) \; \; \; ,
\eqno(8)
$$
$$
{\bar g} = 1 \; + \; {\pi \over 6} \; {h_1 ^2} \, {r^2} \; + \; O ( {r^3} ) \;
\; \; .
\eqno(9)
$$
(Note: equations (8) and (9) correct an error in reference\cite{garfinkle}).  
The
value of the scalar curvature at $r=0$ is given by
$$
R (u,0) = - \, 2 \pi \, {h_1 ^2} \; \; \; .
\eqno(10)
$$

Equation (5) can be regarded as a set of decoupled ODEs for the value of $h$
along each ingoing light ray.  These equations are used to determine the
evolution of $h$  for one time step,
and then the whole process is continued until the scalar field either forms a
black hole or disperses.  The critical solution is found by a binary search of
$p$ parameter space to find the boundary between those data 
that form black holes
and those that do not.  Since $h$ is evolved along ingoing light
rays, the spatial scale of the grid shrinks as the evolution proceeds.  
With the outermost gridpoint
chosen to be the light ray that hits the singularity of the critical
solution, the grid shrinks at the same rate as the spatial features of the
scalar field.  These features are therefore resolved throughout the evolution.

\section{Results}

All runs were done with 300 spatial gridpoints.  The code was run in quadruple
precision on  Dec alpha workstations and in double precision on a Cray YMP8. 
The initial data for the scalar field was chosen to be of the form
$$
\phi (0,r) = p \; {r^2} \; \exp \left [ {{(r - {r_0})}^2}/{\sigma ^2} \right ]
\; \; \; .
\eqno(11)
$$
Here $p$ is our parameter, and $r_0$ and $\sigma $ are constants.  This is the
family which was evolved in \cite{garfinkle}, where the value of 
the critical parameter
$p*$ was found.  Here we evolve this family for 100 values of $p < p*$, chosen
equally spaced in $\ln (p* - p)$.  During each evolution, we keep 
track of the behavior of
the scalar curvature at $r=0$ and thus find the maximum of its 
absolute value $R_{\rm max}$.  

Figure 1
shows a graph of $\ln {R_{\rm max}} $ vs. $\ln (p*-p)$.  
In the figure, each point is the
result of one evolution.  The points were fit to a five parameter 
curve that is a straight line
plus a sine wave.  (Both the figures and the curve fitting were done with
KaleidaGraph).  Figure 1 also shows this curve.  However, 
because of the large number of 
data points and the goodness of the fit, the curve is 
indistinguishable from the data points.

The parameters of the fit are the slope and intercept of the line, 
and the amplitude,
period and phase of the sine wave.  To examine the goodness of the fit, 
the data and the fit 
are plotted in Figure 2 with the straight line piece of the fit 
subtracted from both of them.  
Here, we see that the fit is good, but not exact.  Indeed there 
is no reason for the fit to
be exact: the function should be periodic with period 
$\Delta /(2 \gamma )$ and therefore
a sum of sine waves of period $\Delta /(2 \gamma n)$ for integer $n$.  
In addition, inaccuracies
in the numerical evolution of the spacetime contribute some 
``noise'' to the data points.

Of particular interest are the slope of the line and the period of the 
sine wave.  It is not 
clear what error should be attributed to the parameters of the fit.  
While there is an error
that can be obtained formally from the fitting process, there 
may be additional errors due to
inaccuracies of the numerical evolution algorithm itself.  
Using three significant figures, we 
find that the slope of the line is -0.747 and the period of the 
sine wave is 4.63.  The values 
of $\gamma $ and $\Delta $ given in reference\cite{gundlach1} are
$\gamma = 0.374 \pm 0.001 $ and $\Delta = 3.4453 \pm 0.0005 $.  
These numbers give rise to
$ - \, 2 \gamma = - 0.748 \pm 0.002 $ and 
$ \Delta / (2 \gamma ) = 4.61 \pm 0.01 $.  Thus
it is clear that the slope of the line is $ - 2 \, \gamma $ 
and the period of the sine wave
is $ \Delta / (2 \gamma ) $.  That is, as expected, $R_{\rm max}$ scales like
${(p*-p)}^{- 2 \gamma}$ with a periodic wiggle of period $\Delta / (2 \gamma )$.

\section{Discussion}

Our paper considers the behavior of only one sort of curvature: 
scalar curvature at the
position of the central observer.  Clearly there are other sorts of 
curvature that one
could treat.  The quantities $ {{|{R^{ab}} {R_{ab}}|}^{1/2}}$ and 
$ {{|{R^{abcd}} {R_{abcd}}|}^{1/2}}$ would also be expected to scale 
like ${(p*-p)}^{- 2
\gamma}$.  In the case of spherical symmetry, and evaluated on 
the world line of the central
observer, these quantities yield nothing new.  All of them are 
proportional to $h_1 ^2$. 
This is what one would expect, since in spherical symmetry the 
gravitational field has no
degrees of freedom of its own, and the Ricci tensor just depends 
quadratically on the
gradient of the scalar field.  However, there is no need to restrict 
consideration to the
world line of the central observer.  One could also consider the 
maximum value of the
scalar curvature (or any of these other curvatures) over the 
whole spacetime.  In this
project we chose the world line of the central observer 
mostly for convenience, since it is
very easy to evaluate the scalar curvature there.  We do not expect 
the results to differ
much if instead we find the maximum of the scalar curvature over the whole
spacetime, since in a spherically symmetric collapse we would expect the
spacetime maximum of the scalar curvature to occur at or near the world line
of the central observer.

The situation is different in the case of collapse without 
spherical symmetry.  Here there
is no central observer and so the spacetime maximum of curvature is 
the appropriate
quantity to consider.  (Though in the case of axisymmetry with equatorial
plane reflection symmetry there is a preferred observer).  
Choptuik scaling has been
shown to occur in the collapse of vacuum, axisymmetric gravity waves
\cite{evans2}.  It would be interesting to see whether curvature scaling takes
place in this situation.  Of course, since the spacetimes are vacuum, $R$ and
${R^{ab}} {R_{ab}} $ vanish.  Therefore, the appropriate quantity 
to consider is the
spacetime maximum of $ {{|{R^{abcd}} {R_{abcd}}|}^{1/2}}$.  
We would expect this quantity
to scale like $ {{(p*-p)}^{- 2 \gamma }}$ (with a small periodic 
wiggle) for those
spacetimes that just barely do not collapse to form black holes.  
It would also be
of interest to investigate the gravitational collapse of an axisymmetric scalar
field and look for scaling in the spacetime maxima of $R$ and 
${|{R^{ab}} {R_{ab}}|}^{1/2} $ as well as $ {{|{R^{abcd}} {R_{abcd}}|}^{1/2}}$.

Although we would expect that a quantity with dimensions of length 
would scale as
${{|p-p*|}^\gamma}$, it is known that in some cases this does 
not occur.  Hod and
Piran\cite{piran3} have performed a numerical simulation of the 
collapse of a spherically
symmetric charged scalar field.  Here the black holes formed have 
charge as well as mass. 
Since charge has units of length, one might expect that near 
the critical solution charge
scales as ${{(p-p*)}^\gamma}$.  Instead, the charge vanishes faster:
like ${{(p-p*)}^{2 \gamma}}$.  (This scaling is explained in
references \cite{piran3,g2}). Thus a simple consideration of the dimensions
of a quantity is not sufficient, in all cases, to predict the scaling of that
quantity.  It would be interesting to know which quantities can be expected to
scale as their dimensions would suggest, and which behave 
anomalously. In any case,
some sort of scaling behavior can be expected for many different 
quantities, both
in spacetimes that barely form black holes and in those that barely do not.   
    
\section{Acknowledgements}

This work was partially supported by NSF grant PHY-9722039 and by a Cottrell 
College Science Award of Research Corporation to Oakland University. 
Some of the
computations were performed at the Ohio Supercomputer Center.

\begin{figure}[bth]
\begin{center}
\makebox[4in]{\psfig{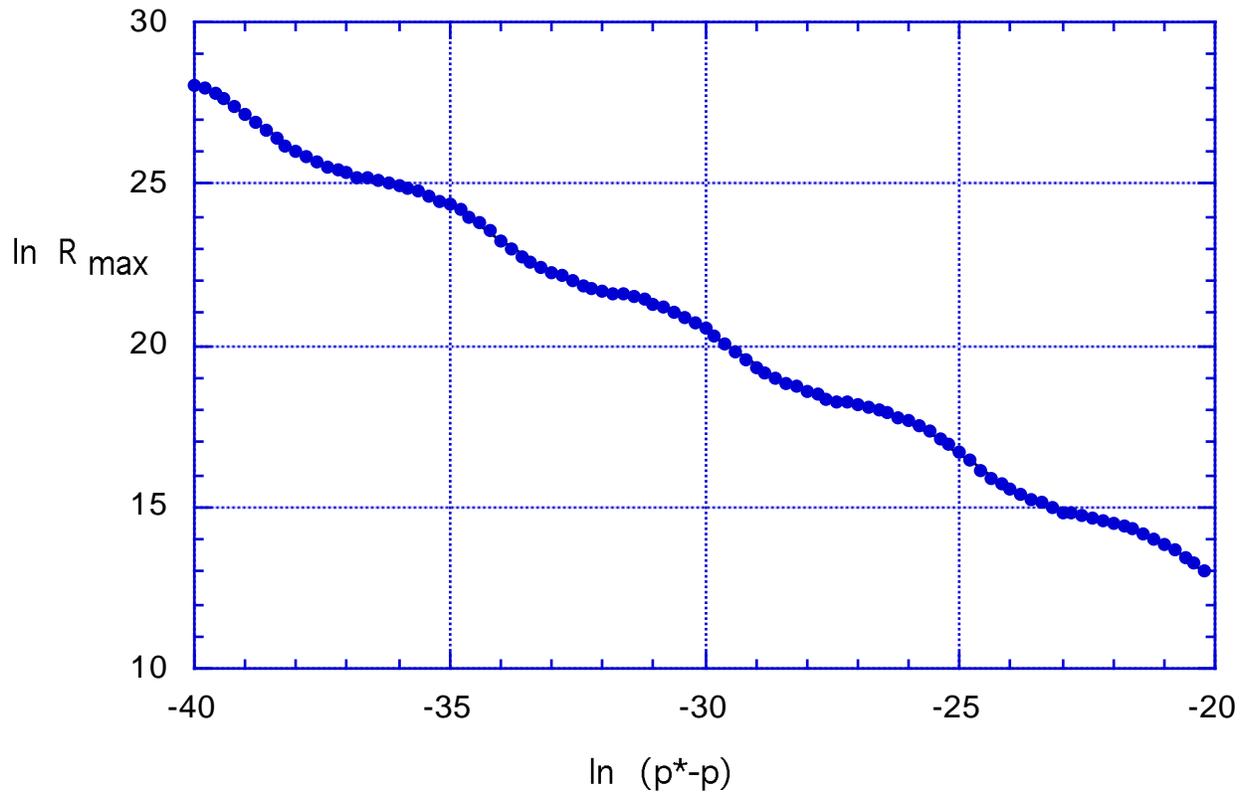}}
\caption{$\ln {R_{\rm max}}$ is plotted vs. $\ln (p*-p)$.  The result is
a line with slope $ - 2 \gamma $ and a periodic wiggle with period $\Delta /(2
\gamma )$}
\label{crvf1}
\end{center}
\end{figure}
\vfill\eject 
\begin{figure}[bth]
\begin{center}
\makebox[4in]{\psfig{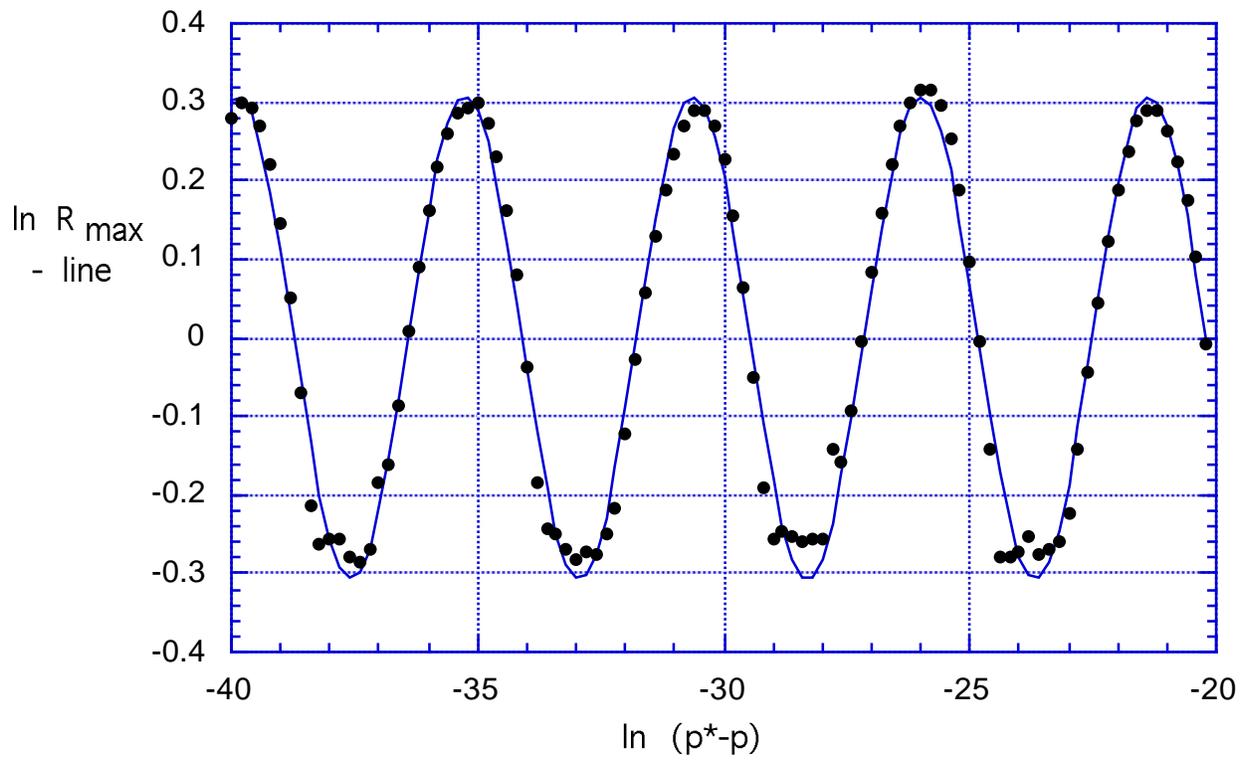}}
\caption{The data and the fitted curve of figure 1 are plotted with the straight
line piece  of the fitted curve removed from both.}
\label{crvf2}
\end{center}
\end{figure}

\end{document}